\documentclass[article]{revtex4-1}
\usepackage[T1]{fontenc}
\usepackage[utf8]{inputenc}
\usepackage{graphicx}
\usepackage{lmodern}
\usepackage{amsfonts}
\usepackage{mathtools}
\usepackage{bm}

\newcommand{\mvec}[1]{\boldsymbol{\mathrm{#1}}}

\begin{document}
\title{Computational applications of the Many Interacting Worlds interpretation of quantum mechanics}
\author{Simone Sturniolo}
\affiliation{Scientific Computing Department, STFC, Rutherford Appleton Laboratory, Harwell Campus, Didcot, OX11 0QX}
\email{simone.sturniolo@stfc.ac.uk}

\begin{abstract}
  While historically many quantum mechanical simulations of molecular dynamics have relied on the Born-Oppenheimer approximation to separate electronic and nuclear behavior, recently a lot of interest has arisen towards quantum effects in nuclear dynamics as well. Due to the computational difficulty of solving the Schr\"{o}dinger equation in full, these effects are often treated with approximate methods.\newline
  In this paper we present a new algorithm to tackle these problems, using an extension to the Many Interacting Worlds approach to quantum mechanics. This technique uses a kernel function to rebuild the probability density and therefore, at a difference with the approximation presented in the original paper, can be naturally extended to $n$-dimensional systems. This opens up the possibility of performing quantum ground state searches with steepest descent methods, and could potentially lead to real time quantum molecular dynamics simulations.\newline
  The behavior of the algorithm is studied in different potentials and numbers of dimensions and compared both to the original approach and to exact Schr\"{o}dinger equation solutions whenever possible.
\end{abstract}

\maketitle

\section{Introduction}  \label{intro}

Since their original introduction \cite{carparr} Ab-Initio Molecular Dynamics have been widely used to study a range of different systems. Historically, these simulations have always relied on the Born-Oppenheimer approximation \cite{bornopp} to separate between electronic and nuclear motions, treating the first with quantum mechanics and the latter with classical Newtonian mechanics. In recent years, however, an interest has arisen towards the relevance of quantum effects in nuclear motions, as advances in computational technology have made their calculation more practical. Many calculations and experiments show that nuclear quantum effects, especially involving the motion of hydrogen nuclei, are relevant to fully describe the behavior of water and ice \cite{quanteff_1, quanteff_2}, in strongly hydrogen bonded systems \cite{quanteff_3} and in biological macromolecules \cite{quanteff_4}. Simulating these effects is no easy feat. One of the most popular approaches is that of Path-Integral Molecular Dynamics, or PIMD \cite{pimd_seminal}, which allows one to approximate quantum statistical distributions by replacing a single nucleus with many copies of it organized as beads in a closed loop, all behaving classically except for a fictitious harmonic potential term linking them together. Since the dynamics of a system defined this way are not necessarily physical any more, different conventions can be adopted for the effective masses of its various vibrational modes depending on the quantities of interest \cite{witt_pimd}, such as Centroid Molecular Dynamics (CMD) \cite{cao_cmd}, Ring Polymer Molecular Dynamics (RPMD) \cite{rpoly_review}, and the one that is most commonly referred as PIMD proper. This technique has been used successfully to explore problems such as the behaviour of hydrogen atoms shared by molecules in water monolayers on metal surfaces \cite{quanteff_1}, the quantum nature of the hydrogen bond \cite{q_hbond} and bimolecular reaction rates \cite{q_rrate}, using both ab-initio methods and parametrized potentials.\newline
In 2014, Hall, Deckert and Wiseman proposed a possible interpretation of quantum mechanics that shares many features with PIMD \cite{miw_seminal}. In this approach, that the authors call Many Interacting Worlds (MIW), quantum mechanical behavior emerges from many copies of the same particle all interacting with each other through a potential that has no classical equivalent. As opposed to PIMD, however, this potential is repulsive, which means MIW could describe a quantum ground state as an equilibrium configuration; in addition, MIW is theoretically an approximation to full quantum dynamics, meaning it should be able to simulate time-dependent quantum evolution. This makes it a promising avenue to explore for the development of new computational techniques for the treatment of nuclear quantum effects. However, while the theory presented in \cite{miw_seminal} is general, the practical implementation proposed in the paper can be applied only to 1-dimensional systems. In this paper we develop a different approach that is naturally extensible to higher dimensions and could therefore be put to practical use in molecular dynamics simulations.

\section{The Many Interacting Worlds approach} \label{miw_approach}

\subsection{Theory}\label{miw_theory}

The MIW approach, as presented in \cite{miw_seminal}, can be considered as a discretization of the Holland-Poirier hydrodynamical approach to QM \cite{holland, schiff_poirier} or it can be derived from the well known deBroglie-Bohm pilot wave interpretation \cite{bohm1952suggested1, bohm1952suggested2}. For the full derivation, we direct the reader to the original paper. Here we just give an outline of the fundamentals of this approach.\newline
The system to be described is represented by a number of worlds $N$, with a multi-world configuration at time $t$

\begin{equation}
  \label{miw_config}
  \boldsymbol{\mathrm{X}}(t) = \left\{ \boldsymbol{\mathrm{x}}_1(t), \boldsymbol{\mathrm{x}}_2(t), \cdots, \boldsymbol{\mathrm{x}}_N(t) \right\}
\end{equation}

with every $\boldsymbol{\mathrm{x}}_n(t) = \left[x_{1,n}(t), x_{2,n}(t), \cdots, x_{K,n}(t)\right]$ being the total classical configuration of world $n$; namely, an array of the $K$ degrees of freedom of the system. For a generic $D$-dimensional system containing $Q$ particles it will be $K=QD$. It is easy to see how then the probability density to find the system in a configuration $\boldsymbol{\mathrm{q}}$, equivalent to the square modulus of the wave function in the usual Schr\"{o}dinger's picture, can be approximated as

\begin{equation}
  \label{miw_prob_dens}
  P(\boldsymbol{\mathrm{q}}, t) = \left|\Psi(\boldsymbol{\mathrm{q}}, t)\right|^2 \sim  \sum_{n=1}^N{\delta(\boldsymbol{\mathrm{q}} - \boldsymbol{\mathrm{x}}_n(t))}
\end{equation}

using the Dirac delta distribution. The dynamics of the system are governed by the usual laws of Newtonian mechanics. The classical Hamiltonian can be written as:

\begin{equation}
  \label{miw_hamil}
  \mathbb{H}_{MIW}\left(\boldsymbol{\mathrm{X}}\right) = \sum_{n=1}^N{\left[\sum_{k=1}^K{\frac{1}{2}m_k\dot{x}_{k,n}^2} + V(\boldsymbol{\mathrm{x}}_n)\right]} + U_{MW}(\boldsymbol{\mathrm{X}})
\end{equation}

where the quantities with index $k$ (masses, coordinates etc.) correspond to each individual particle, and the potential is a function of the entire world's configuration. One can distinguish a term which operates on each world configuration separately (with the classical potential $V$ also including any regular interactions among particles, like electrostatic forces) and an inter-world potential $U_{MW}$, which is non-classical in nature and introduces quantum effects. For example, delocalization is the consequence of $U_{MW}$ being repulsive and preventing all particles to find an equilibrium in the potential minimum, and energy indeterminacy is the consequence of energy being exchanged between worlds thanks to the inter-world coupling and therefore not being conserved in each separate world (while the overall many-world ensemble is indeed conservative).\newline
The general form of $U_{MW}$ is:

\begin{equation}
  \label{miw_potential}
  U_{MW}(\boldsymbol{\mathrm{X}}) = \sum_{n=1}^N\sum_{k=1}^K\frac{1}{2m_k} \left[g_N^k(\boldsymbol{\mathrm{x}}_n; \boldsymbol{\mathrm{X}})\right]^2
\end{equation}

where

\begin{equation}
  \label{miw_gfunc}
  g_N^k(\boldsymbol{\mathrm{q}}; \boldsymbol{\mathrm{X}}) \approx \frac{\hbar}{2}\frac{1}{P(\boldsymbol{\mathrm{q}}; \boldsymbol{\mathrm{X}})}\frac{\partial P(\boldsymbol{\mathrm{q}}; \boldsymbol{\mathrm{X}})}{\partial q_k}
\end{equation}

Here $P$ represents the distribution describing the probability to find the system in a given configuration $\boldsymbol{\mathrm{q}}$, as in equation \ref{miw_prob_dens}, but its dependence from the configuration of the `world particle' $\boldsymbol{\mathrm{X}}$ is made explicit.\newline
From Eq. \ref{miw_potential} and \ref{miw_gfunc} it is clear that to run a simulation based on the MIW approach it is necessary to rebuild the probability density function $P(\boldsymbol{\mathrm{q}}; \boldsymbol{\mathrm{X}})$ in some approximated way for a given multi world configuration. This is implied in the choice of only writing an "approximate" equality in Eq. \ref{miw_gfunc}. Eq. \ref{miw_prob_dens} suggests one way to do this, but it is obvious that in practical computation, where limits on available power and time will force one to use a small number of worlds $N$, this method would fail rather badly. In \cite{miw_seminal}, the authors propose for the 1D case of a single particle an approximation

\begin{equation}
  \label{miw_basic_pdens}
  P(x_n) \approx \frac{1}{N(x_{n+1}-x_n)}
\end{equation}

Here $x_n$ has become a scalar, since each world has only one degree of freedom. This equation holds whenever the distance between the same particle in adjacent worlds is slowly varying and by enforcing that $x_{n+1}>x_n$ all the time, and does indeed produce good results. This leads to an inter-world potential depending overall on five worlds - the world of interest $n$ and its first and second neighbors. It has however two problems that prevent it from being applicable to general purpose simulations, namely that it can not be naturally extended to more than one dimension and that it features a divergent potential which makes numerical integration very sensitive to the time step used when any two world-particles happen to be close enough.\newline
In this paper we suggest a different method to compute the probability density which overcomes these problems, using the technique known as kernel density estimation (KDE) \cite{rosenblatt1956, parzen1962}. The idea is simply to apply a kernel distribution $\mathcal{K}$ to Eq. \ref{miw_prob_dens}, so that

\begin{equation}
\label{miw_kernel_pdens}
P(\mvec{q}) =  \sum_{n=1}^N{\mathcal{K}(\mvec{q}-\mvec{x}_n)}
\end{equation}

In this way, and with a good choice of function $\mathcal{K}$, $P(\mvec{q})$ is continuous and differentiable on all space, which leads to a natural way of computing the quantum forces, and can be defined similarly for any dimensionality. A very similar approach has been proposed in parallel to this work by Herrmann and the authors of the original MIW paper in \cite{miw_kde_gstate}. In that, a more natural multi-dimensional extension of the original method by using Delaunay triangulations is explored as well, but it is found to be inconvenient for practical applications due to the discontinuities it introduces in the dynamics.\newline
When using KDE, an obvious choice is to make the kernel function Gaussian, which, including the necessary normalization conditions, returns:

\begin{equation}
  \label{miw_gkernel_pdens}
  P(\mvec{q}) =  \frac{1}{N(\sqrt{\pi}b)^D}\sum_{n=1}^N{\exp\left[-\frac{(\mvec{q}-\mvec{x}_n)^2}{b^2}\right]}
\end{equation}

with $b$ a free kernel bandwidth parameter and $D$ number of dimensions of the system. It is then possible to derive analytically the potential and the forces. While the process is not especially complex, the calculations are long, and are reported in Appendix \ref{app:ker_forces}.\newline
The Gaussian kernel however has the potential to give rise to a problem. Let us consider the case of a simulation of a single quantum particle. From now on, it must be clear that when we talk about "particles" we mean in fact multiple classical copies of the same particle across worlds interacting only through the interworld potential, and not effectively different particles interacting classically. Due to the appearance of the derivative of $P$ in Eq. \ref{miw_gfunc}, one can see after deriving the forces that it will give rise to no repulsion when two particles are close enough or overlapping. This runs counter physical intuition: since the interworld potential must reproduce the effects of what we could call `quantumness' on the system, it should be generally repulsive, to avoid the wave function collapsing in a single spot and losing position indeterminacy. This is a property of any symmetric and smooth kernel, as its derivative in the center will always be null. Therefore, if the particles happened to get closer than a certain distance during the simulation, they might end up coalescing and this artifact would compromise the final result. For this reason we test also a different kernel, with a discontinuous, non-zero derivative in the origin:

\begin{equation}
\label{miw_ekernel_pdens}
P(\mvec{q}) =  \frac{\Gamma(D/2)}{2N(D-1)!(\sqrt{\pi}b)^D}\sum_{n=1}^N{\exp\left[-\frac{|\mvec{q}-\mvec{x}_n|}{b}\right]}
\end{equation}
 
where the proper normalization factor has been inserted in front (with $\Gamma$ meaning the gamma function). Since this factor is less obvious, proof of how it's derived is provided in Appendix \ref{app:expker_norm}. Figure \ref{fig:kerProb} compares the $U_{MW}$ for two particles as the distance between them varies for both kernels and highlights the problem and the way the exponential kernel solves it. Potential and forces can be found for this kernel similarly to what has been seen with the Gaussian one, and are written out in Appendix \ref{app:ker_forces} as well.\newline

\begin{figure}
  \includegraphics[width=0.7\textwidth]{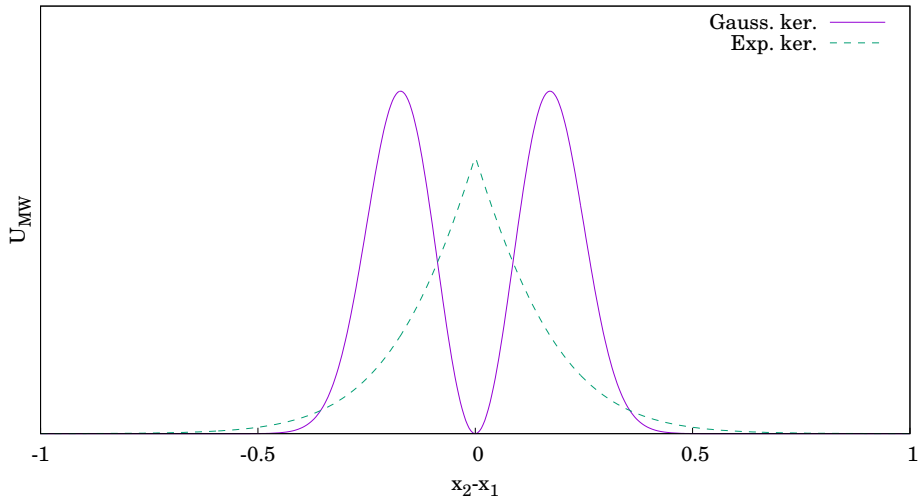}
  \caption{Many world potential for $N=2$ and for the cases of Gaussian and exponential kernels. It can be seen how the former features a minimum for the case of overlapping particles where the latter has a cusp. Units are arbitrary.}
  \label{fig:kerProb}
\end{figure}

In section \ref{miw_simdetails} we will proceed to test the MIW method in some numerical simulations on toy models and compare its results with both solutions obtained by traditional methods based on diagonalization of the Hamiltonian and, for the case of 1D problems, MIW simulations carried out with the potential derived from Eq. \ref{miw_basic_pdens}.

\subsection{Comparison with methods of the PIMD family}

Given the similarities between the two techniques, it is worth the effort to explore a bit more the analogies and differences between MIW and PIMD-like methods (PIMD, CMD and RPMD), to better evaluate the potential applicability of this new approach. As a note, from now on, I will refer to all these latter three methods as PIMD unless specified otherwise.\newline
As mentioned already, MIW and PIMD share a fundamental similarity in their approach to simulating quantum effect, as both use a number of classical simulations coupled by a fictitious potential to reproduce quantum statistics. The potential that couples PIMD beads, however, is harmonic and works only between next neighbour worlds, which in turn allows to treat the dynamics by separating independent harmonic modes. This is not possible in kernel MIW, where the potential is strongly non-linear and couples all worlds with all others. Nevertheless, the formal similarities means that for a lot of existing software packages implementing MIW dynamics could probably be a relatively easy task, as it could reuse much of the PIMD code.\newline
Performance wise, these two methods tend to be complementary and that makes it hard to set up a direct comparison. A key feature of PIMD methods is that the number of beads required to converge a calculation increases dramatically as the temperature approaches 0 K \cite{pimd_tdep}. Therefore, PIMD methods tend to perform better at higher temperatures. Conversely, the theory behind MIW simulations justifies best their use for searching the ground state, and therefore the low temperature limit. In addition, the harmonic potential featuring in PIMD is attractive; the equilibrium configuration for the beads would be one where they all sit in the same potential minimum, and therefore the dynamical simulation is vital to actually sample quantum statistics. The MIW potential on the other hand is repulsive, and its equilibrium should correspond to an approximation of the quantum ground state. This means that it should be possible to find ground state densities using not only molecular dynamics, but even common optimisation algorithms such as BFGS \cite{fletcher1987practical}. While, as seen in the next section, MIW calculations seem to require slightly higher numbers of worlds than a typical room-temperature PIMD simulation (for which 16 or 32 beads are common values), the increased calculation load can be amply compensated by replacing a costly and long molecular dynamics simulation with a simple geometry optimisation. This strategy is explored in the next section.\newline
Finally, there's the issue of real-time quantum dynamics. While PIMD technically computes an evolution in imaginary time, and thus can't be directly interpreted as a dynamical process, it is possible with both CMD and RPMD to compute quantum time correlation functions \cite{witt_pimd, q_rtimecorr}. Ideally, MIW simulations should be able to achieve a similar result in a more immediate way, as they represent evolution in real time instead. However, the accuracy of this evolution is limited by how well the quality of the reconstructed density is preserved, and errors accumulating through time will probably cause time correlation functions to be accurate only on a short time scale.

\section{Simulation details}\label{miw_simdetails}

Numerical simulations were carried out on a personal computer using Python and the scientific libraries Numpy and Scipy for matrix diagonalizations and optimization operations \cite{numpy, scipy}. 
\newline
When necessary, the exact solution results in 1D were obtained by building an Hamiltonian based on a matrix Numerov method \cite{mnumerov}. This approach was then expanded to higher dimensionality; the details are explained in Appendix \ref{app:numerov}. Since this method uses a direct space basis set, all potentials are treated effectively as if they were enclosed in an infinite well. Whenever harmonic potentials appear, the known analytical solutions for ground state energies and wavefunctions are used.\newline
For the MIW method, the equations of motion were integrated using a standard velocity Verlet algorithm, and a Langevin thermostat was used for thermalization when required. In addition, an adaptive time step has been used, where at any given step $i$:

\begin{equation}\label{adaptive_dt}
dt_i = min\left(dt_0\frac{max(|F_0|)}{max(|F_i|)}, dt_{max}\right)
\end{equation}

so that $dt$ scales with the maximum force present in the system.\newline
Particular attention, of course, must be paid to the initial estimate of the bandwidth parameter, $b$, which controls the radius of the interaction. This is a common problem in kernel density estimation, well known in statistics. Given the particle positions, one has to find the kernel width that best fits the target probability distribution. For cases where the initial desired probability distribution is known (for example, when initialising a simulation with knowledge of the ground state), the AMISE method was used. For those where instead only an educated guess was possible, the Silverman method was employed. Both these methods are described in \cite{kde_review}.

\section{Results}

\subsection{Energy}

In these tests we focus on the performance of the MIW approach in dealing with the ground state of a proton in a few example potentials. We initialise the MIW system by using the known ground state probability density obtained from diagonalising the Hamiltonian and consider the error in the energy so obtained, in order to compare the different kernels and dimensionalities in ideal conditions.\newline
The potentials used are of two types. One is a simple harmonic potential of the form 

\begin{equation}\label{harm_potdef}
V_{harm}(\mvec{x}) = \frac{k}{2}\mvec{x}^2
\end{equation}
 
while the other is a multi-dimensional Lennard-Jones potential with an angular term of the form

\begin{equation}\label{lj_potdef}
V_{lj}(\mvec{x}) = \Delta V_{r}\left[\left(\frac{|\mvec{x}-\mvec{x}_0|}{r_0}\right)^{-12}-2\left(\frac{|\mvec{x}-\mvec{x}_0|}{r_0}\right)^{-6}\right] + \Delta V_{\alpha}\left(1-\frac{\mvec{x}-\mvec{x}_0}{|\mvec{x}-\mvec{x}_0|}\cdot\hat{i}\right)
\end{equation}

where $\hat{i}$ is the versor of the x-axis. This was chosen to represent a crude approximation of a chemical bond. In this part we use three such potentials, which from now on we will label \texttt{harm1}, \texttt{harm10} and \texttt{lj1}. The first two are defined by Equation \ref{harm_potdef} with $k=1$ and $k=10\,\mathrm{eV/\r{A}}^2$ respectively. The third uses Equation \ref{lj_potdef} with $\Delta V_r=1\,\mathrm{eV}$, $\mvec{x}_0=-2.5\hat{i}\,\mathrm{\r{A}}$, $r_0=1\,\mathrm{\r{A}}$ and $\Delta V_\alpha=10\,\mathrm{eV}$. It should be noted that the origin for the LJ potential was chosen because all simulations were ran in box-shaped grids ranging from -2 to 2 \AA. The grids had 200, 40 and 15 points of side respectively for 1, 2 and 3D. The wave function is considered zero outside of this space. Thus, this choice allows to have the minimum of the potential inside the box without including the singularity, which could cause problems.\newline
The initial particle positions were generated in two different ways. The first method was to distribute the particles so that each grid element contains just the right amount to match the target distribution as closely as possible. The second instead employed a simple Monte Carlo method to randomly distribute them, following the target distribution but allowing for random fluctuations.

\begin{figure*}[h]
  \includegraphics[width=\textwidth]{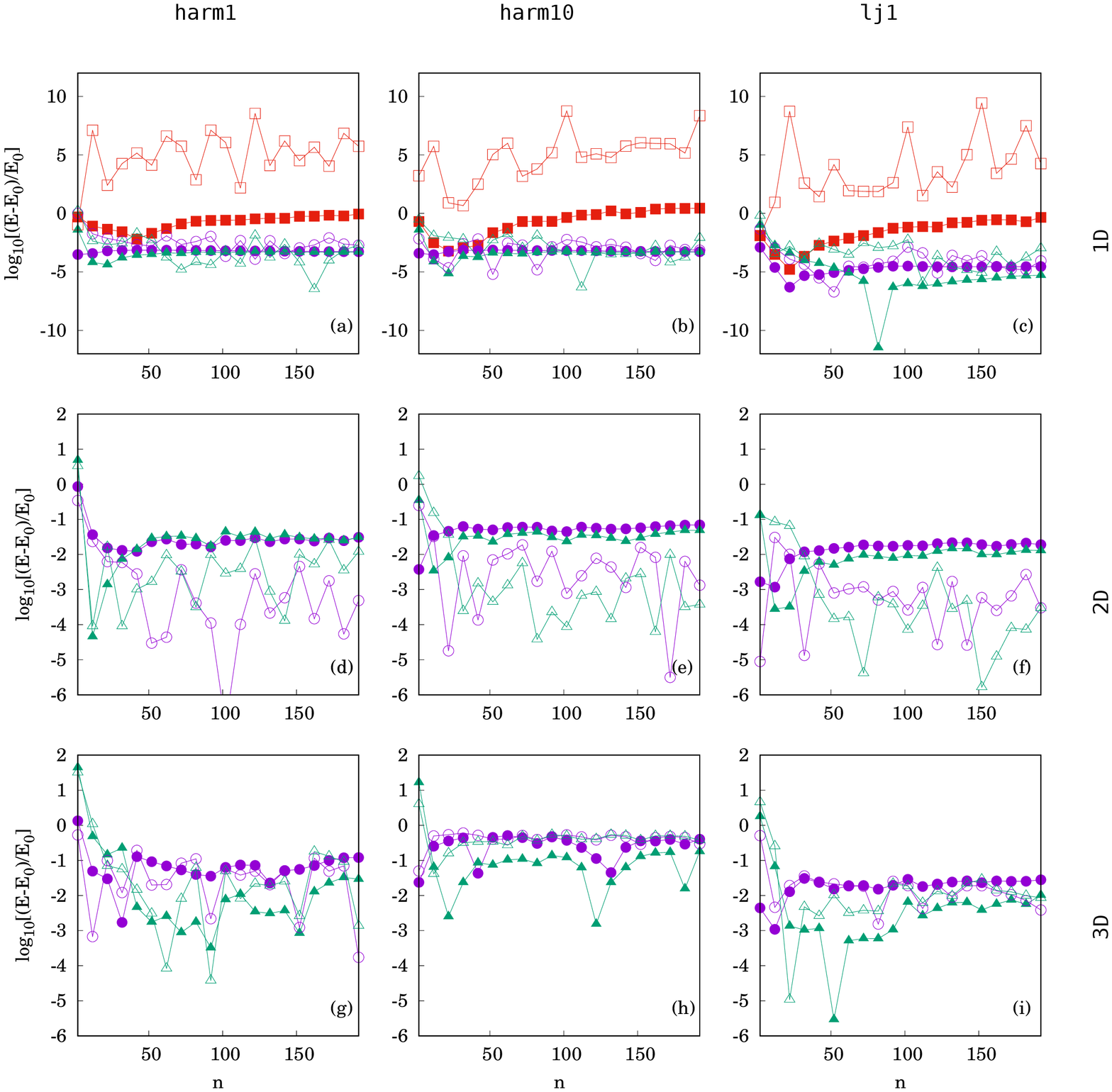}
  \caption{Logarithmic plots of the MIW energy error with an ideal distribution vs. number of worlds for various potentials and dimensionalities. Ideal energies were computed with a matrix Numerov algorithm using grids of 200, 40 and 15 points of side respectively for 1, 2 and 3D. Circles represent the Gaussian kernel, triangles the exponential one, and squares the method from the original paper (only applicable to 1D). Filled dots represent uniformly distributed particles, empty ones the Monte-Carlo distributed ones. Legend: (a) (b) (c) are 1D potentials, (d) (e) (f) are 2D, (g) (h) (i) are 3D. By potential type, (a) (d) (g) use the \texttt{harm1} potential, (b) (e) (h) use \texttt{harm10}, (c) (f) (i) use \texttt{lj1}. The same labels apply to figures \ref{fig:groundstate_E} and \ref{fig:gstate_chi}.}
  \label{fig:EvsN}
\end{figure*}

When calculating the energy within the MIW approximation in order to compare it to the energy found by diagonalising the Hamiltonian, care needs to be taken. In the basic approach, forces are calculated on the hypothesis of perfectly point-like particles, and would in fact be exact for an infinite number of particles with infinitesimally small spacing. When using kernels, each particle contributes to the overall density with a distributed density. This can be interpreted as each particle representing, in fact, a large or infinite number of particles distributed according to that function and moving around rigidly. This brings forth two main consequences for energy calculations:

\begin{itemize}
  \item the $i$-th particle's contribution to potential energy, $V_i$, should in theory not be calculated in a pointlike manner, $V(x_i)$, but rather as the integral
  \begin{equation}
    V_i = \int_{\mathbb{R}^D}{V(\mvec{x})\mathcal{K}(\mvec{x})d\mvec{x}}
  \end{equation}
  This might not always be possible in actual calculations. In that case, the point-like approximation is accurate to the first order, since the kernel is symmetric. If one has access to the second derivative of the potential it is possible to expand it in Taylor series and find a third order approximation which also depends on $b$;
  
 \item there is an "internal energy" correction for each particle, consisting of the many world interaction energy of the particles constituting the Gaussian distribution kernel itself. This is a constant term and can be calculated by applying Equation \ref{miw_potential}, replacing the probability density in Eq. \ref{miw_gfunc} with the kernel function and the sum over $n$ with an integral over all space. Luckily, it is rather easy to calculate for both kernels:
  \begin{equation}\label{kernel_E}
    U_{corr}^{(gauss)} = \frac{\hbar^2}{4m}\frac{D}{b^2} \qquad U_{corr}^{(exp)} = \frac{\hbar^2}{8m}\frac{D}{b^2}
  \end{equation}
  It should be remarked that this term is not required when comparing two MIW simulations with the same parameters, being constant; it becomes necessary however if considering simulations of different kind or with different $b$.
\end{itemize}

The total energy is therefore computed as:

\begin{equation}\label{totE}
E_{tot} = U_{MW}(\mvec{X})+U_{corr}^{(kernel)}+\sum_i^N\left(V_i+\frac{1}{2}mv_i^2\right)
\end{equation}

though effectively the kinetic energy term at the end is zero for non-dynamical calculations like this one and the search for the ground state presented in the next subsection.\newline
Figure \ref{fig:EvsN} shows the error in energy calculated with the MIW approach for the various potentials and dimensionalities tested. A few observations are in order. Convergence is overall satisfactory in all cases, with various degrees of success. The 1D case shows obviously the advantages of the kernel approximation compared to the one given in Equation \ref{miw_basic_pdens}. For the kernel approximation, both with Gaussian and exponential kernels, increasing the number of worlds used tends to reliably improve convergence, which eventually reaches a limit value. As a general rule it seems that for these systems using any more than 50 worlds does not really bring any improvement in the approximation of the energy. The simpler method, on the other hand, converges only initially to then immediately diverge again when the density of worlds becomes too high, as its dependency on the inverse of the distance between world-particles makes it far more sensitive to numerical errors. This is even more obvious for Monte-Carlo generated particle configurations. This means that there is a non-trivial optimal amount of worlds to use, that in any real world application would be another variable to consider when deciding the parameters for a calculation.\newline
In the 2D and 3D cases, random Monte-Carlo initialization provides a better average approximation but also greater noise, whereas uniform distributions quickly converge to a slightly biased value. This is probably the effect of such distributions being dependent on the underlying grid, which introduces artefacts.\newline

\subsection{Ground state convergence}
Now we move on to investigating a method of finding the ground state of a potential by using the MIW approach. This is straightforward: we generate a system of a number of worlds (in all cases here, $N=50$ was used) in some configuration that we consider a reasonable starting point, then we use some optimisation routine to converge it down to a point where all forces are in equilibrium. Here we try two different approaches to this process. In addition to that, since it's possible that particles might get stuck in non-physical configurations or local minima, the simulation is periodically re-initialized by computing the density and using it to re-distribute the particles. This was done using the uniform distribution method, which is found to give the better results. When a re-initialization is performed, the bandwidth is newly calculated too, using the AMISE method.\newline
Here two methods were used. The first is a simple molecular dynamics simulation, with a strong damping achieved by using a Langevin thermostat with $T = 0\,K$ and $\gamma=10^{15}s^{-1}$. For this simulation, 10 sequences of 1000 time steps, with $dt_{max} = 5\cdot10^{-17}s$ ($3\cdot10^{-17}s$ for the 3D case) were used, with one re-initialization between each sequence. The second uses the Scipy implementation of the BFGS optimization algorithm, using 10 sequences of a maximum of 40 iterations, with a tolerance of $1\cdot10^{-5}eV/\AA$ on the forces.
The initial configuration was chosen to be a completely uniform distribution for the harmonic potentials and a Gaussian centred on the minimum for the Lennard-Jones ones for the 1D and 2D cases. This choice was made because the latter, being much flatter on the long distance, risked causing convergence problems to an ensemble of particles that is too spread out. In the 3D case, a Gaussian of arbitrary width was used for all three potentials. This was not considered a problem as it seems reasonable to expect that in all practical applications similar assumptions could be made, and the classical minimum of the potential would likely be known from previous simulations.

\begin{figure*}[h]
  \includegraphics[width=\textwidth]{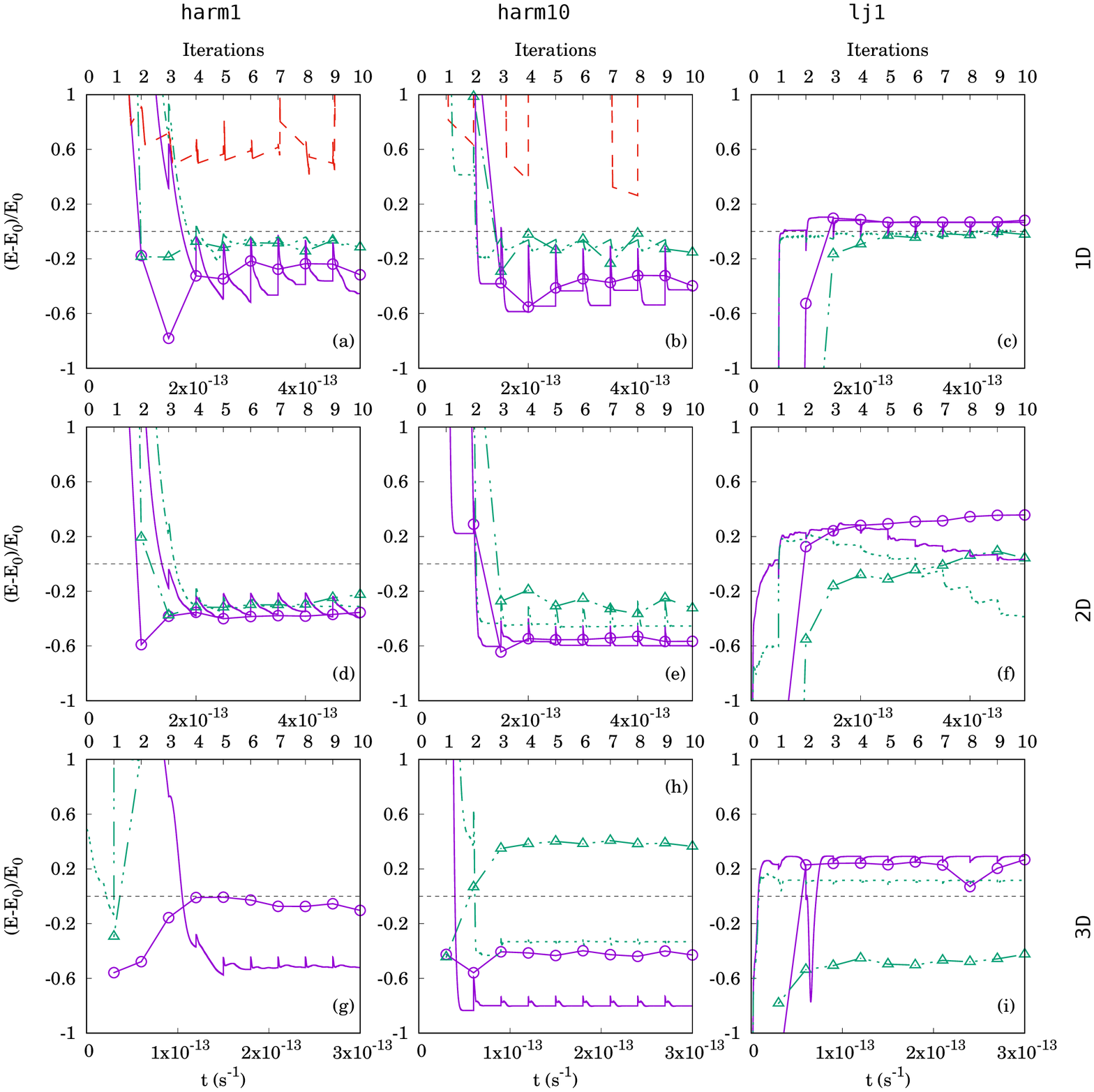}
  \caption{Energy convergence during the relaxation process for different potentials and dimensionalities. Continuous lines represent the Gaussian kernel, dot-dashed lines the exponential one, and dashed lines the method from \cite{miw_seminal} converged with damped Langevin MD. Empty circles and triangles represent respectively the Gaussian and exponential kernels converged with the BFGS algorithm. For the labels, refer to the caption in figure \ref{fig:EvsN}.}
\label{fig:groundstate_E}
\end{figure*}

\begin{figure*}[h]
	\includegraphics[width=\textwidth]{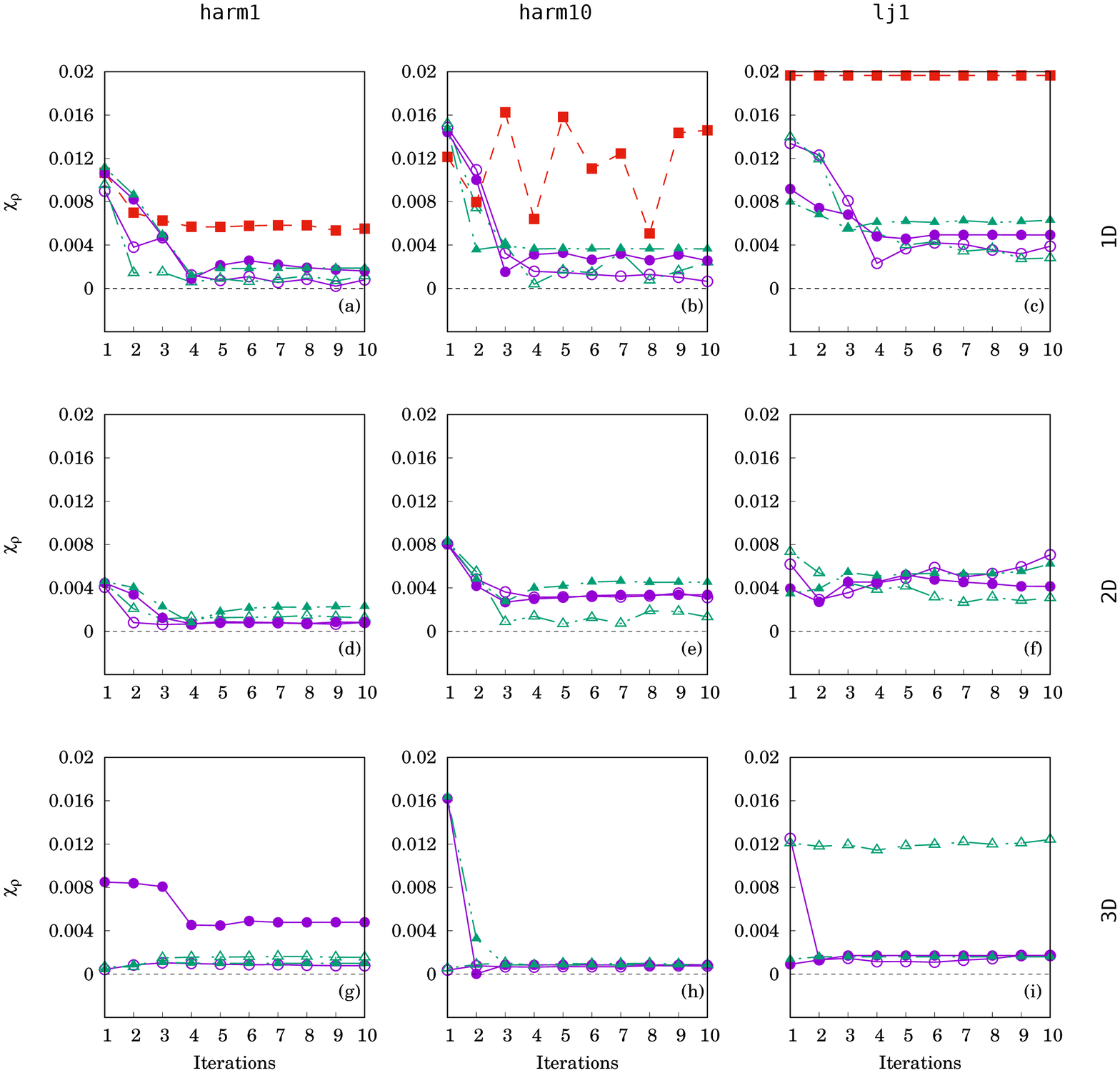}
	\caption{Ground state density error RSS convergence during the relaxation process for different potentials and dimensionalities. Circles represent the Gaussian kernel and triangles the exponential one. Full markers represent damped MD, whereas empty ones represent BFGS. It was not possible to compute the quantity for the original method as it does not provide a continuous approximation for the density. For the labels, refer to the caption in figure \ref{fig:EvsN}.}
	\label{fig:gstate_chi}
\end{figure*}

The results of the calculation can be seen in Figures \ref{fig:groundstate_E} and \ref{fig:gstate_chi}, for convergence of energy over time and the root sum square (RSS) of the probability density error on the grid, that we label as $\chi$. For energies obtained with the damped MD method, all points throughout all iterations are provided. For every other quantity, only the final values of each of the 10 iterations is recorded instead. Energies and densities were compared to the known analytical solutions for harmonic potentials, and to numerically computed solutions for the Lennard-Jones one.\newline
For damped MD, the energy error clearly displays dents corresponding to each reinitialization, but after the first few iterations it generally falls back to its converged value. As a general rule, one can see the exponential kernel performing generally slightly better, except for the \texttt{lj1} 2D case, where the energy seems to diverge. For BFGS the behaviour is often rather similar, with a few exceptions (for example the energy and density of the 3D \texttt{lj1} case with the exponential kernel have a bigger error than any other approach).\newline
In terms of performance, the damped MD method requires by definition 10000 evaluations for both the energy and the forces of the entire system. By comparison, the BFGS runs required a number of energy and forces evaluations both approximately comprised between 600 and 1800. From these results, one can see how the latter seems definitely much more computationally convenient, while producing results that are comparable to damped MD. This would be a unique advantage of this approach over the existing path-integral based techniques, which require a full MD run to produce results even for the ground state.\newline
In general, this approach to ground state search appears to be promising but potentially sensitive to the choices of parameters made. Periodically reinitializing the configuration or other corrective approaches can be used to prevent it from developing artefacts.

\subsection{Finite temperature effects}

We now move on to examining a simple example of possible application of MIW simulations to the realm of finite temperature quantum dynamics. While the original MIW theory does not explicitly mention temperature, there is no reason to think that it should not be possible to simulate incoherent finite temperature quantum dynamics by simply plugging one of the well known MD thermostats into a MIW simulation. This is a consequence of the fact that thermostats approximate the system's interaction with the environment, and classical interactions between different particles in a MIW simulation are perfectly equivalent to the ones in a regular simulation. An interesting question is whether the thermostats should be correlated or coupled across worlds. Intuitively, correlated thermostats would represent an environment that is concentrated in a relatively small region of the phase space and evolves coherently in time, whereas uncorrelated thermostats would represent an environment widely dispersed in phase space and decohered. While there may be some interesting insights to be gained from exploring this matter, for the time being we will settle for fully uncorrelated thermostats, that seem to paint a much more realistic portrait of the situation, especially for high temperatures.\newline

\begin{figure}[h]
	\includegraphics[width=\textwidth]{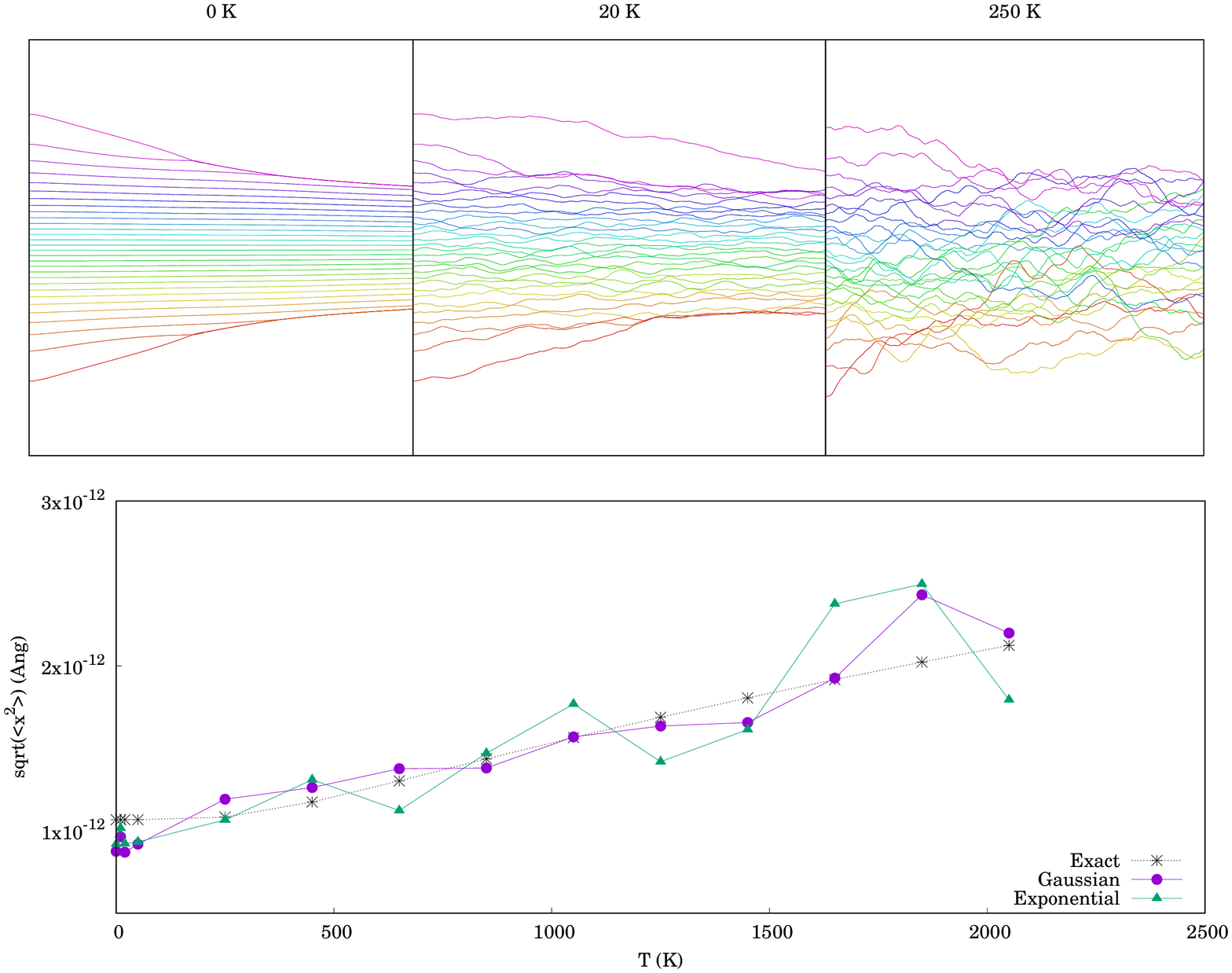}
	\caption{Top: Many Interacting World trajectories at different temperatures and computed value of $\sqrt{<x^2>}$ up to $2000\,K$ for a particle in a harmonic oscillator with $k=1\,\mathrm{eV/\mathring{A}^2}$, using $N=30$ MIW worlds and an exponential kernel.\newline
		Bottom: standard deviation for theoretical and computed densities with Gaussian and exponential kernels as a function of temperature.}
	\label{fig:thermal}
\end{figure}

Figure \ref{fig:thermal} gives us a simple insight in how MIW simulations can reproduce thermal effects. The full bundle of trajectories, from starting configuration to the end of a molecular dynamics simulation, are shown for three different temperatures. The $0\,K$ case is a perfect example of a damped MIW system converging to its ground state, with the contraction (driven by the external potential) being eventually countered by the repulsion due to the MIW potential, finding an equilibrium. It should be remarked however that while these are an approximation to "true" Bohmian trajectories, they are affected by the limits of the method. Specifically, at this temperature an artefact can be seen since at the fringes of the configuration the particles tend to "bunch up" instead of spreading more as they should. This tends to happen even more when using a Gaussian kernel, which is affected by the problems described in section \ref{miw_theory}. At the higher temperatures, the trajectories get scrambled and the system expands, which reduces the importance of this coalescence effect as well. The collisions may transfer further energy among particles so that fluctuations will be bigger than they would be in a non-interacting ensemble, and may allow particles to overcome barriers that should be impassable (thus allowing tunnelling). Ultimately, when the temperature is high enough, the MIW potential's contribution becomes tiny compared to the thermostat forces, and the system reaches the classical limit.

\begin{figure}[h]
	\includegraphics[width=0.7\textwidth]{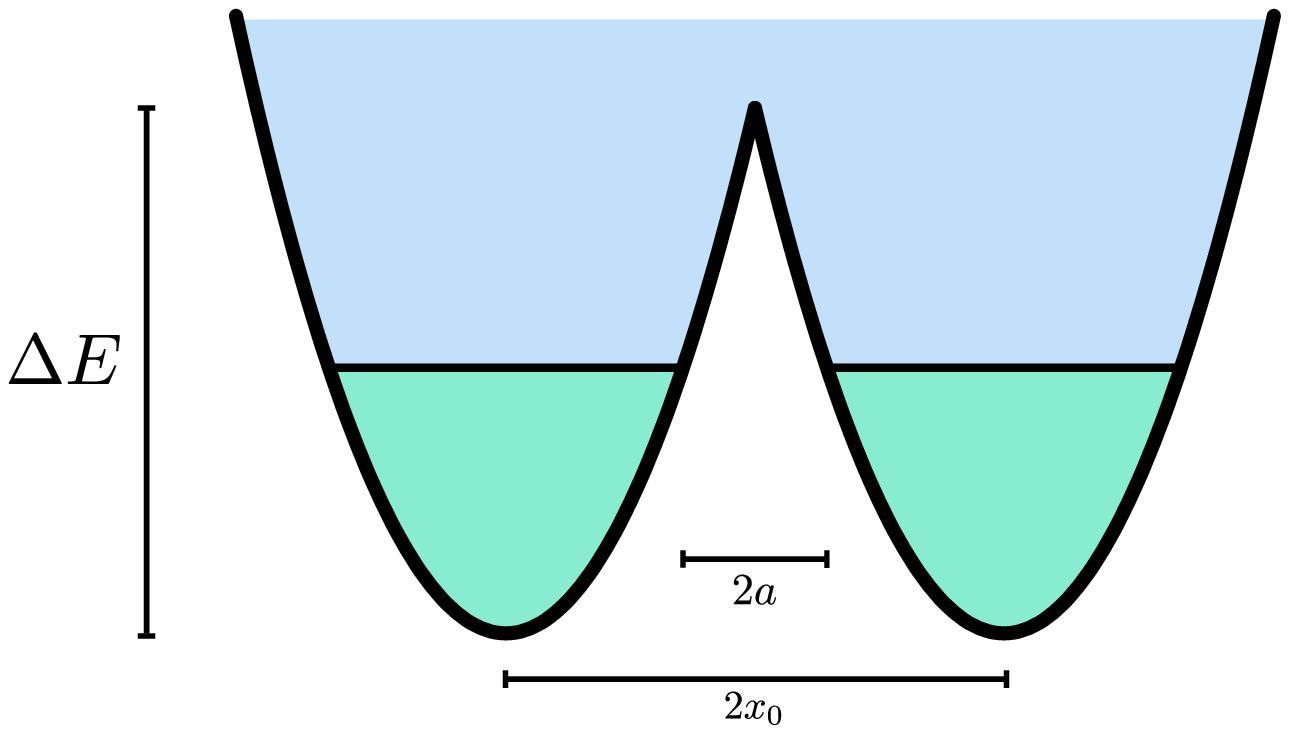}
	\caption{Double harmonic well potential as described by R.P. Bell in \cite{bell_tunnel}. The barrier height is $\Delta E$, $x_0$ is the distance of the minimum from the barrier and $a$ is the distance of the turning point - namely, the point where the potential exceeds the zero point energy of the particle.}
	\label{fig:dwell}
\end{figure}

The system chosen for testing whether the MIW approach can reproduce temperature-dependent quantum tunnelling rates is a simple double well built by joining two harmonic potentials along a plane, as seen in Figure \ref{fig:dwell}. This system has been studied by Bell \cite{bell_tunnel} and its tunnelling rate temperature dependence is known. We choose a potential formed by two harmonic wells of $k=10\,\mathrm{eV/\mathring{A}^2}$
with $x_0=0.2\,\mathrm{\mathring{A}}$. This leads to a potential barrier of $\Delta E = kx_0^2/2 = 0.2\,\mathrm{eV}$ and turning points situated at $a\sim0.06\,\mathrm{\mathring{A}}$. The Arrhenius classical jumping rate is:

\begin{equation}\label{arr_jump}
\nu_{c} = \nu_0\exp\left(-\frac{\Delta E}{k_BT}\right)
\end{equation} 

whereas the Bell quantum corrected version is

\begin{equation}\label{bell_jump}
\nu_{q} = \nu_0\frac{1}{\beta-\frac{\Delta E}{k_BT}}\left[\beta\exp\left(-\frac{\Delta E}{k_BT}\right)
-\frac{\Delta E}{k_BT}\exp(-\beta)\right]
\end{equation} 

with

\begin{equation}\label{bell_beta}
\beta =  \frac{a \pi\sqrt{2m \Delta E}}{\hbar}\sim 1.77
\end{equation}

Three separate simulations were run with $N=50$ worlds: one with a Gaussian kernel, one with an exponential kernel, and one with no kernel forces at all, making it effectively 50 decoupled classical simulations. A Langevin thermostat with $\gamma=10^{14}\,s^{-1}$ was used. Here a little digression is in order. It is common wisdom that Langevin thermostats should not be used when computing diffusion rates; however, there's reason to believe this is justified in this specific case. The rationale for not using it in ordinary MD simulations is that a Langevin thermostat fully couples each individual particle to the heat bath, and this is unrealistic for, for example, molecules in a fluid. However this is not the case here: we are effectively simulating only one particle, and each copy we do simulate is in fact fully coupled, classically, to its own heat bath, namely, the rest of its world. There is no doubt, of course, that the chosen $\gamma$ will control the time scale of the process (in fact, it seems hardly a coincidence that as seen latter we will find $\nu_0=\gamma$). However, since we are interested in comparing jumping rates, and how the MIW potential enhances them, rather than in their absolute values, this is not necessarily a problem. For multi-particle simulations of course the usual considerations would apply, and a Nos\'{e}-Hoover thermostat would be more suited to the task at hand.\newline
Tunnelling was calculated by measuring the fraction of the density inside the starting well and fitting an exponential decay curve to it as it fell from its initial value of almost 1 (some leakage due to the tails of the distributions is present) to the equilibrium value of 0.5. The no kernel simulation was used as benchmark to fit the value of $\nu_0$, using Equation \ref{arr_jump}, which was then plugged into Equation \ref{bell_jump} to estimate the quantum jumping rate.

\begin{figure}[h]
	\includegraphics[width=0.7\textwidth]{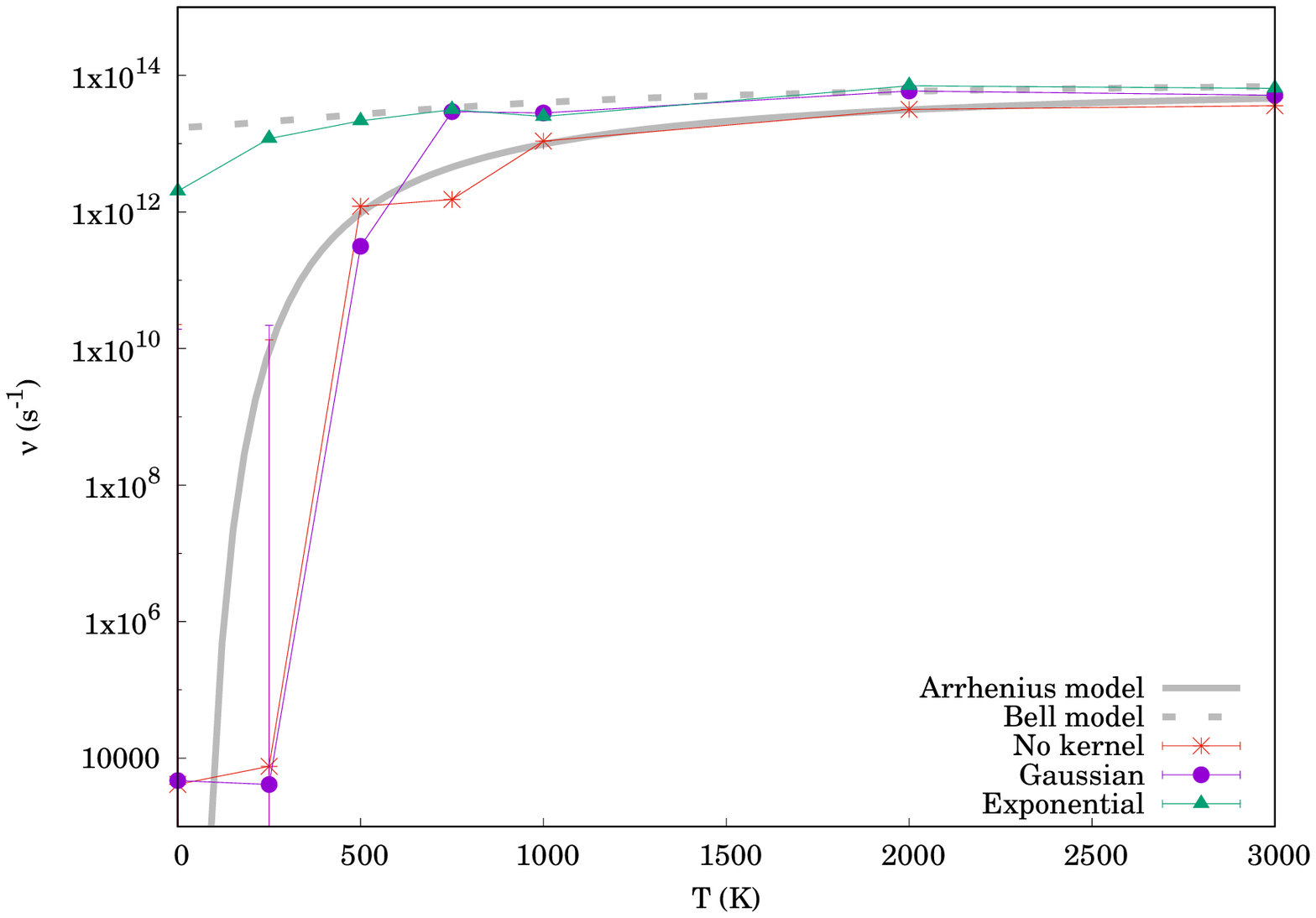}
	\caption{Jumping rates in a MIW simulation on a double harmonic well potential. The fitted parameters are shown with error bars (though most of them are so small as to be invisible) and overlapped with the Arrhenius and Bell models.}
	\label{fig:tunnel}
\end{figure}

Figure \ref{fig:tunnel} shows the final result. The Arrhenius model was fitted with $\nu_0=10^{14}\,s^{-1}$, and the result applied to the Bell model. The rates originated from the exponential kernel simulations follow it closely, showing that the quantum MIW potential does indeed enhance the jumping process and reproduce the correct tunnelling dynamics. The Gaussian kernel simulations behave closer to the ones without kernel at the beginning and then catch up with the quantum model around $T=750\,K$. This is probably due to the already mentioned issue with particles `bonding' when using a smooth kernel, and the problem is overcome once the system has enough kinetic energy to break those pairs. The calculation suggests that it is indeed possible to simulate finite temperature quantum dynamical effects with the MIW method.

\section{Conclusions and future prospects}

An extension to the Many Interacting Worlds description of QM first introduced in \cite{miw_seminal} that makes use of kernel density estimation has been proposed. The method appears to give promising results in reproducing the solutions of simple quantum problems with an ensemble of coupled classical simulations and opens up a novel road to real time finite temperature quantum dynamics for ab-initio molecular dynamics and the study of quantum nuclear effects.\newline
Some details of the method need to be better understood before applying it to molecular dynamics problems. For example, the Gaussian kernel has been shown to often perform worse than the exponential one due to its smoothness; however, the same quality makes it ideal to approximate the true density distribution. A way to overcome the smoothness problem would be desirable. A possible road would be to make the kernel width $b$ a dynamical variable, allowing kernels to squeeze when two particles come too close so that their repulsion grows and they never get to the point of overlapping. This would make the calculations more complex but would also add more degrees of freedom to the system and possibly make it better at approximating the true wave function. A recent work \cite{stein_miw} also suggests a connection between the choice made for probability density reconstruction and which quantum state the particles effectively approximate.  The logic is hard to translate to the kernel method used here, but if possible it might shed some light on a way to simulate excited states specifically. The work done in \cite{miw_kde_gstate} shows how this can effectively work in 1D, provided that the positions of the nodes of the wavefunction are known beforehand.\newline Finally, the analogies between the MIW and the PIMD methods are striking and suggest that a deeper connection between the two might exist. Studying that might bring new insights on how to mitigate each method's weaknesses by mixing it with the other.

\begin{acknowledgments}
Thanks for the useful discussions to Phil Hasnip, Leonardo Bernasconi and Dominik Jochym. This work was conducted within the framework of the CCP for NMR crystallography, which is funded by the EPSRC grants EP/J010510/1 and EP/M022501/1.
\end{acknowledgments}

\appendix

\section{Calculation of quantum forces with Gaussian and exponential kernels}\label{app:ker_forces}

We now show how equation \ref{miw_potential} and its derivatives can be computed efficiently assuming the probability is constructed with a Gaussian kernel, as seen in \ref{miw_kernel_pdens}. This formalism is designed to make for especially compact code when working with languages that allow for element-wise array operations like Fortran or Python+Numpy. Let us consider the case of a single quantum particle represented with $N$ worlds in $D$ dimensions. The coordinate of the particle in world $i$ along dimension $k$ is written as $x_i^{(k)}$. Let us also consider a kernel of fixed bandwidth $b$. Then we define:

\begin{align} \label{kernel_symbols_1}
 &r_{ij}^{(k)} = x_i^{(k)}-x_j^{(k)}\qquad r_{ij}^2 = \sum_k (r_{ij}^{(k)})^2 \qquad P_{ij} = \frac{1}{N(\sqrt{\pi}b)^D}exp\left(-\frac{ r_{ij}^2}{b^2}\right) \\
 &P_{ij}^{\prime(k)} = -\frac{2}{b^2}r_{ij}^{(k)}P_{ij} \qquad P_{ij}^{\prime\prime} = -\frac{2}{b^2}(1-\frac{2}{b^2}r_{ij}^2)P_{ij} \nonumber
\end{align}

In this formalism, $P_{ij}$ represents the contribution of particle $j$ to the density probability at the position of particle $i$, $P_{ij}^{\prime}$ its gradient with respect to the position of particle $i$, and so on. It should be noted that, while we included the normalisation factor in Eq. \ref{kernel_symbols_1}, this is not really relevant for the final forces as it disappears in the formulation of $U_{MW} $, which contains a ratio between the kernel and its derivative.\newline
Computing the total quantities, summed over all particles, requires a bit more care. If we define:

\begin{equation}\label{kernel_symbols_2}
P_i = \sum_jP_{ij}
\end{equation}

as the total density probability at the position of particle $i$, then its derivatives are:

\begin{align}\label{kernel_P_d1}
\frac{dP_i}{dx_n^{(l)}} =
\begin{dcases}
P_n^{\prime(l)} & n=i\\
-P_{in}^{\prime(l)} & n\neq i
\end{dcases}
\end{align}

and the second ones:

\begin{align}\label{kernel_P_d2}
\frac{dP_i^{\prime(k)}}{dx_n^{(l)}} =
\begin{dcases}
P_n^{\prime\prime(l)} & n=i, l=k\\
-P_{in}^{\prime\prime(l)} & n\neq i, l=k\\
-\frac{2}{b^2}\sum_j r_{nj}^{(k)}P_{nj}^{\prime(l)} & n = i, l \neq k \\
\frac{2}{b^2}r_{in}^{(k)}P_{in}^{\prime(l)} & n \neq i, l\neq k
\end{dcases}
\end{align}

where the quantities with only one index ($P_n', P_n^{\prime\prime}$) represent sums over $j$ as seen in Eq. \ref{kernel_symbols_2}.
Using these relationships, it's straightforward, if rather tedious, to compute the interworld potential and the forces. We can rewrite eq. \ref{miw_potential} for this kernel using the new formalism:

\begin{equation}\label{kernel_g_U}
g_i^{(k)} = \frac{\hbar}{2}\frac{P_i^{\prime(k)}}{P_i} \qquad
U = \frac{1}{2m}\sum_{i,k}\left[g_i^{(k)}\right]^2
\end{equation}

Then the full derivative with respect to the particle positions can be written as:

\begin{align}\label{kernel_F}
\frac{dU}{dx_n^{(l)}} &= 2g_n^{(l)}\left[-\frac{1}{P_n^2}\left(P_n^{\prime(l)}\right)^2+\frac{1}{P_n}P_n^{\prime\prime(l)}\right] + \\
&\sum_{i\neq n}2g_i^{(l)}\left[\frac{1}{P_i^2}P_{in}^{\prime(l)}P_i^{\prime(l)}-\frac{1}{P_i}P_{in}^{\prime\prime(l)}\right] + \nonumber\\
&\sum_{k \neq l}2g_n^{(k)}\left[-\frac{1}{P_n^2}P_n^{\prime(k)}P_n^{\prime(l)}-\frac{2}{b^2P_n}\sum_jr_{nj}^{(k)}P_{nj}^{\prime(l)}\right] + \nonumber\\
&\sum_{i \neq n, k \neq l}2g_i^{(k)}\left[\frac{1}{P_i^2}P_{in}^{\prime(k)}P_i^{\prime(l)}+\frac{2}{b^2P_i}r_{in}^{(k)}P_{in}^{\prime(l)}\right] \nonumber
\end{align}

where it should be noted that the two bottom summation terms are always going to be zero in the 1-dimensional case, which therefore noticeably simplifies the expression. The forces of course are going to be equal to this expression with a minus sign.\newline
Now we consider the case of an exponential kernel. A lot of the passages are similar, but we need to take into account the different form of the derivatives. In this case we have:

\begin{align} \label{ekernel_symbols_1}
&r_{ij} = \sqrt{\sum_k(r_{ij}^{(k)})^2} \qquad P_{ij} = \frac{\Gamma(D/2)}{2N(D-1)!(\sqrt{\pi}b)^D}exp\left(-\frac{r_{ij}}{b}\right) \\
&P_{ij}^{\prime(k)} = -\frac{1}{b}\frac{r_{ij}^{(k)}}{r_{ij}}P_{ij} \qquad P_{ij}^{\prime\prime} = -\frac{1}{b}\frac{1}{r_{ij}}\left[1-\frac{(r_{ij}^{(k)})^2}{r_{ij}^2}-\frac{1}{b}\frac{(r_{ij}^{(k)})^2}{r_{ij}}\right]P_{ij} \nonumber
\end{align}

With these new assignments, Eq. \ref{kernel_P_d1} still holds for first derivatives, whereas second derivatives become:

\begin{align}\label{ekernel_P_d2}
	\frac{dP_i^{\prime(k)}}{dx_n^{(l)}} =
	\begin{dcases}
		P_n^{\prime\prime(l)} & n=i, l=k\\
		-P_{in}^{\prime\prime(l)} & n\neq i, l=k\\
		-\sum_j \frac{r_{nj}^{(k)}}{r_{nj}}\left(\frac{1}{r_{nj}}+\frac{1}{b}\right)P_{nj}^{\prime(l)} & n = i, l \neq k \\
		\frac{r_{in}^{(k)}}{r_{in}}\left(\frac{1}{r_{in}}+\frac{1}{b}\right)P_{in}^{\prime(l)}& n \neq i, l\neq k
	\end{dcases}
\end{align}

and therefore the forces:

\begin{align}\label{ekernel_F}
	\frac{dU}{dx_n^{(l)}} &= 2g_n^{(l)}\left[-\frac{1}{P_n^2}\left(P_n^{\prime(l)}\right)^2+\frac{1}{P_n}P_n^{\prime\prime(l)}\right] + \\
	&\sum_{i\neq n}2g_i^{(l)}\left[\frac{1}{P_i^2}P_{in}^{\prime(l)}P_i^{\prime(l)}-\frac{1}{P_i}P_{in}^{\prime\prime(l)}\right] + \nonumber\\
	&\sum_{k \neq l}2g_n^{(k)}\left[-\frac{1}{P_n^2}P_n^{\prime(k)}P_n^{\prime(l)}-\frac{1}{P_n}\sum_j\frac{r_{nj}^{(k)}}{r_{nj}}\left(\frac{1}{r_{nj}}+\frac{1}{b}\right)P_{nj}^{\prime(l)}\right] + \nonumber\\
	&\sum_{i \neq n, k \neq l}2g_i^{(k)}\left[\frac{1}{P_i^2}P_{in}^{\prime(k)}P_i^{\prime(l)}+\frac{1}{P_i}\frac{r_{in}^{(k)}}{r_{in}}\left(\frac{1}{r_{in}}+\frac{1}{b}\right)P_{in}^{\prime(l)}\right] \nonumber
\end{align}

\section{Normalization of the exponential kernel}\label{app:expker_norm}

The multivariate exponential kernel centered in the origin is defined as:

\begin{equation}
\mathcal{K}(\mvec{q}) = \exp\left[-\frac{|\mvec{q}|}{b}\right]
\end{equation}

for a given width $b$.\newline
This needs to be divided by its integral over the entire space for normalization purposes. For the 1D case the solution is simple, as the integral 

\begin{equation}
\int_0^\infty{\exp\left(-\frac{x}{b}\right) dx} = b
\end{equation}

is easily found, and thus the overall integral is $2b$. However the multivariate case is more complex. One can find it considering two things. First, the integral of a radial function in a $D-$dimensional space can be defined as

\begin{equation}
\int_{\mathbb{R}^n}f(|\mvec{q}|)d\mvec{q} = \int_0^\infty{f(r)\omega_{D-1}(r)dr}
\end{equation}

where $\omega_{D-1}$ is the surface area of the $D-$dimensional sphere \cite{stromberg1981introduction}. This is known to be

\begin{equation}
\omega_{D-1}(r) = \frac{2\pi^{\frac{D}{2}}}{\Gamma\left(\frac{D}{2}\right)}r^{D-1}
\end{equation}

On the other hand, the radial integral can be carried out by parts if we notice that

\begin{align}
\int_0^\infty{\exp\left(-\frac{r}{b}\right)r^{D-1}dr} &= \left|-b\exp\left(-\frac{r}{b}\right)r^{D-1}\right|_0^\infty\\
&+ (D-1)b\int_0^\infty{\exp\left(-\frac{r}{b}\right)r^{D-2}dr}\\
&= (D-1)b\int_0^\infty{\exp\left(-\frac{r}{b}\right)r^{D-2}dr}
\end{align}

as the first term goes to zero both on $r=0$ and $r=\infty$. We can repeat the operation $D-1$ times, thus finding:

\begin{equation}
\int_0^\infty{\exp\left(-\frac{r}{b}\right)r^{D-1}dr} = (D-1)!b^D
\end{equation}

which combined to the prefactor for the surface area of an n-sphere gives us

\begin{equation}
\int_{\mathbb{R}^D}  \mathcal{K}(\mvec{q}) = \frac{2(\sqrt{\pi}b)^D}{\Gamma\left(\frac{D}{2}\right)} (D-1)!
\end{equation}

whose reciprocal is the normalization factor we need.

\section{A matrix Numerov method for integration of the Schr\"{o}dinger equation in arbitrary dimensions}\label{app:numerov}

The original 1D matrix Numerov method for integrating the Schr\"{o}dinger equation was presented in \cite{mnumerov}. An analogue scheme for the 2D equation is described in \cite{MOHEBBI}. Here we write the same scheme in a general form for any number of dimensions.\newline
Similarly to what happens in 1D, the Numerov method is designed to solve equations of the form

\begin{equation}
\nabla^2\psi(\mvec{x}) = f(\mvec{x})\psi(\mvec{x})
\end{equation}

where, in the case of the Schr\"{o}dinger equation, 

\begin{equation}
f(\mvec{x}) = -\frac{2m}{\hbar^2}(E-V(\mvec{x}))
\end{equation}

Now we expand $\psi$ in a Taylor series, define the function on a grid, and consider the `stencil' surrounding a grid point composed by all nearest neighbors - the points that are one step forward or backward in each direction. Then we can write

\begin{equation}
\sum_i^D\frac{\psi(\mvec{x}+h_i\mvec{\epsilon}_i)-2\psi(\mvec{x})+\psi(\mvec{x}-h_i\mvec{\epsilon}_i)}{h_i^2} = f\psi + \frac{1}{12}\sum_i^D\frac{\partial^4\psi}{\partial x_i^4}h_i^2 + \mathcal{O}(h^6)
\end{equation}

with $\mvec{\epsilon}_i$ unit vector and $h_i$ grid step for dimension $i$. In this case the relation holds

\begin{equation}
\nabla^2(f\psi) = \nabla^2(\nabla^2\psi) = \sum_i^D \frac{\partial^4\psi}{\partial x_i^4} + \sum_i^D\sum_{j\neq i}^D \frac{\partial^4\psi}{\partial x_i^2\partial x_j^2} 
\end{equation}

so we can separate

\begin{equation}
\begin{split}
 \sum_i^D\frac{\psi(\mvec{x}+h_i\mvec{\epsilon}_i)-2\psi(\mvec{x})+\psi(\mvec{x}-h_i\mvec{\epsilon}_i)}{h_i^2} & = f\psi + \frac{1}{12}\sum_i^D\frac{\partial^2(f\psi)}{\partial x_i^2}h_i^2 \\
&-\frac{1}{12}\sum_i^Dh_i^2\sum_{j\neq i}^D \frac{\partial^4\psi}{\partial x_i^2\partial x_j^2}
\end{split}
\end{equation}

Now, considering that we are working within a grid of finite size, we can write all operators as matrices. The matrix $\mathbf{A}$ as described in \cite{mnumerov} extends to a Kronecker sum:

\begin{equation}
\mathbf{A}^{(D)} = \sum_i^D \mathbf{A}_i^{(D)} =
\bigoplus_i^D \frac{(\mathbb{I}_{-1}-2\mathbb{I}_{0}+\mathbb{I}_{1})}{h_i^2}
\end{equation}

while the matrix $\mathbf{B}$, which operates on $f\psi$ on the right-hand side, becomes:

\begin{equation}
\mathbf{B}^{(D)} = \mathbb{I} + \frac{1}{12}\sum_i^D h_i^2\mathbf{A}_i^{(D)}
\end{equation}

We can also write the mixed derivatives as matrix products

\begin{equation}
\frac{\partial^4}{\partial x_i^2 \partial x_j^2} \rightarrow \mathbf{A}_i^{(D)}\mathbf{A}_j^{(D)}
\end{equation}

which happen to commute since the $\mathbf{A}$ matrices are symmetric. So in the end we can write the multi-dimensional equivalent of the 1D Numerov method as:

\begin{equation}
\begin{split}
& -\frac{\hbar^2}{2m}\left[\mathbf{A}^{(D)}+\frac{1}{12}\sum_{i,j>i}\mathbf{A}_i^{(D)}\mathbf{A}_j^{(D)}(h_i^2+h_j^2)\right]\psi = \mathbf{B}^{(D)}(E-\mathbf{V})\psi \implies \\
& -\frac{\hbar^2}{2m}\mathbf{B}^{-1(D)}\left[\mathbf{A}^{(D)}+\frac{1}{12}\sum_{i,j>i}\mathbf{A}_i^{(D)}\mathbf{A}_j^{(D)}(h_i^2+h_j^2)\right]\psi +\mathbf{V}\psi= E\psi 
\end{split}
\end{equation}

with $\mathbf{V}$ a matrix having the potential along its diagonal and zero everywhere else.
Therefore, this becomes an eigenvalue problem that can be solved by diagonalizing the matrix:

\begin{equation}
\mathbf{M} = -\frac{\hbar^2}{2m}\mathbf{B}^{-1(D)}\left[\mathbf{A}^{(D)}+\frac{1}{12}\sum_{i,j>i}\mathbf{A}_i^{(D)}\mathbf{A}_j^{(D)}(h_i^2+h_j^2)\right] +\mathbf{V}
\end{equation}

and will give us energies and eigenstates as a result.


\begin{thebibliography}{30}
	\providecommand{\natexlab}[1]{#1}
	\providecommand{\url}[1]{\texttt{#1}}
	\expandafter\ifx\csname urlstyle\endcsname\relax
	\providecommand{\doi}[1]{doi: #1}\else
	\providecommand{\doi}{doi: \begingroup \urlstyle{rm}\Url}\fi
	
	\bibitem[Car and Parrinello(1985)]{carparr}
	R.~Car and M.~Parrinello.
	\newblock Unified approach for molecular dynamics and density-functional
	theory.
	\newblock \emph{Phys. Rev. Lett.}, 55:\penalty0 2471--2474, Nov 1985.
	\newblock \doi{10.1103/PhysRevLett.55.2471}.
	\newblock URL \url{http://link.aps.org/doi/10.1103/PhysRevLett.55.2471}.
	
	\bibitem[Born and Oppenheimer(1927)]{bornopp}
	M.~Born and R.~Oppenheimer.
	\newblock Zur quantentheorie der molekeln.
	\newblock \emph{Annalen der Physik}, 389\penalty0 (20):\penalty0 457--484,
	1927.
	\newblock ISSN 1521-3889.
	\newblock \doi{10.1002/andp.19273892002}.
	\newblock URL \url{http://dx.doi.org/10.1002/andp.19273892002}.
	
	\bibitem[Li et~al.(2010)Li, Probert, Alavi, and Michaelides]{quanteff_1}
	Xin-Zheng Li, Matthew I.~J. Probert, Ali Alavi, and Angelos Michaelides.
	\newblock Quantum nature of the proton in water-hydroxyl overlayers on metal
	surfaces.
	\newblock \emph{Phys. Rev. Lett.}, 104:\penalty0 066102, Feb 2010.
	\newblock \doi{10.1103/PhysRevLett.104.066102}.
	\newblock URL \url{http://link.aps.org/doi/10.1103/PhysRevLett.104.066102}.
	
	\bibitem[Pamuk et~al.(2012)Pamuk, Soler, Ram\'irez, Herrero, Stephens, Allen,
	and Fern\'andez-Serra]{quanteff_2}
	B.~Pamuk, J.~M. Soler, R.~Ram\'irez, C.~P. Herrero, P.~W. Stephens, P.~B.
	Allen, and M.-V. Fern\'andez-Serra.
	\newblock Anomalous nuclear quantum effects in ice.
	\newblock \emph{Phys. Rev. Lett.}, 108:\penalty0 193003, May 2012.
	\newblock \doi{10.1103/PhysRevLett.108.193003}.
	\newblock URL \url{http://link.aps.org/doi/10.1103/PhysRevLett.108.193003}.
	
	\bibitem[Wikfeldt(2014)]{quanteff_3}
	K~T Wikfeldt.
	\newblock Nuclear quantum effects in a 1-d model of hydrogen bonded
	ferroelectrics.
	\newblock \emph{Journal of Physics: Conference Series}, 571\penalty0
	(1):\penalty0 012012, 2014.
	\newblock URL \url{http://stacks.iop.org/1742-6596/571/i=1/a=012012}.
	
	\bibitem[Nagel and Klinman(2006)]{quanteff_4}
	Zachary~D. Nagel and Judith~P. Klinman.
	\newblock Tunneling and dynamics in enzymatic hydride transfer.
	\newblock \emph{Chemical Reviews}, 106\penalty0 (8):\penalty0 3095--3118, 2006.
	\newblock \doi{10.1021/cr050301x}.
	\newblock URL \url{http://dx.doi.org/10.1021/cr050301x}.
	\newblock PMID: 16895320.
	
	\bibitem[Marx and Parrinello(1994)]{pimd_seminal}
	Dominik Marx and Michele Parrinello.
	\newblock Ab initio path-integral molecular dynamics.
	\newblock \emph{Zeitschrift f\"ur Physik B Condensed Matter}, 95\penalty0
	(2):\penalty0 143--144, 1994.
	\newblock ISSN 0722-3277.
	\newblock \doi{10.1007/BF01312185}.
	\newblock URL \url{http://dx.doi.org/10.1007/BF01312185}.
	
	\bibitem[Witt et~al.(2009)Witt, Ivanov, Shiga, Forbert, and Marx]{witt_pimd}
	Alexander Witt, Sergei~D. Ivanov, Motoyuki Shiga, Harald Forbert, and Dominik
	Marx.
	\newblock On the applicability of centroid and ring polymer path integral
	molecular dynamics for vibrational spectroscopy.
	\newblock \emph{The Journal of Chemical Physics}, 130\penalty0 (19):\penalty0
	194510, 2009.
	\newblock \doi{10.1063/1.3125009}.
	\newblock URL \url{https://doi.org/10.1063/1.3125009}.
	
	\bibitem[Cao and Voth(1993)]{cao_cmd}
	Jianshu Cao and Gregory~A. Voth.
	\newblock A new perspective on quantum time correlation functions.
	\newblock \emph{The Journal of Chemical Physics}, 99\penalty0 (12):\penalty0
	10070--10073, 1993.
	\newblock \doi{10.1063/1.465512}.
	\newblock URL \url{https://doi.org/10.1063/1.465512}.
	
	\bibitem[Habershon et~al.(2013)Habershon, Manolopoulos, Markland, and
	III]{rpoly_review}
	Scott Habershon, David~E. Manolopoulos, Thomas~E. Markland, and Thomas~F.{\
	}Miller III.
	\newblock Ring-polymer molecular dynamics: Quantum effects in chemical dynamics
	from classical trajectories in an extended phase space.
	\newblock \emph{Annual Review of Physical Chemistry}, 64\penalty0 (1):\penalty0
	387--413, 2013.
	\newblock \doi{10.1146/annurev-physchem-040412-110122}.
	\newblock URL \url{https://doi.org/10.1146/annurev-physchem-040412-110122}.
	\newblock PMID: 23298242.
	
	\bibitem[Li et~al.(2011)Li, Walker, and Michaelides]{q_hbond}
	Xin-Zheng Li, Brent Walker, and Angelos Michaelides.
	\newblock Quantum nature of the hydrogen bond.
	\newblock \emph{Proc. Natl. Acad. Sci. USA}, 108\penalty0 (16):\penalty0
	6369--6373, 2011.
	
	\bibitem[Suleimanov et~al.(2011)Suleimanov, Collepardo-Guevara, and
	Manolopoulos]{q_rrate}
	Yury~V. Suleimanov, Rosana Collepardo-Guevara, and David~E. Manolopoulos.
	\newblock Bimolecular reaction rates from ring polymer molecular dynamics:
	Application to {H + CH4 > H2 + CH3}.
	\newblock \emph{The Journal of Chemical Physics}, 134\penalty0 (4):\penalty0
	044131, 2011.
	\newblock \doi{10.1063/1.3533275}.
	\newblock URL \url{https://doi.org/10.1063/1.3533275}.
	
	\bibitem[Hall et~al.(2014)Hall, Deckert, and Wiseman]{miw_seminal}
	Michael~J.W. Hall, Dirk-Andr\'e Deckert, and Howard~M. Wiseman.
	\newblock Quantum phenomena modeled by interactions between many classical
	worlds.
	\newblock \emph{Phys. Rev. X}, 4:\penalty0 041013, Oct 2014.
	\newblock \doi{10.1103/PhysRevX.4.041013}.
	\newblock URL \url{http://link.aps.org/doi/10.1103/PhysRevX.4.041013}.
	
	\bibitem[Holland(2005)]{holland}
	Peter Holland.
	\newblock Computing the wavefunction from trajectories: particle and wave
	pictures in quantum mechanics and their relation.
	\newblock \emph{Annals of Physics}, 315\penalty0 (2):\penalty0 505 -- 531,
	2005.
	\newblock ISSN 0003-4916.
	\newblock \doi{http://dx.doi.org/10.1016/j.aop.2004.09.008}.
	\newblock URL
	\url{http://www.sciencedirect.com/science/article/pii/S0003491604001757}.
	
	\bibitem[Schiff and Poirier(2012)]{schiff_poirier}
	Jeremy Schiff and Bill Poirier.
	\newblock Communication: Quantum mechanics without wavefunctions.
	\newblock \emph{The Journal of Chemical Physics}, 136\penalty0 (3):\penalty0
	031102, 2012.
	\newblock \doi{http://dx.doi.org/10.1063/1.3680558}.
	\newblock URL
	\url{http://scitation.aip.org/content/aip/journal/jcp/136/3/10.1063/1.3680558}.
	
	\bibitem[Bohm(1952{\natexlab{a}})]{bohm1952suggested1}
	David Bohm.
	\newblock A suggested interpretation of the quantum theory in terms of" hidden"
	variables. i.
	\newblock \emph{Physical Review}, 85\penalty0 (2):\penalty0 166,
	1952{\natexlab{a}}.
	
	\bibitem[Bohm(1952{\natexlab{b}})]{bohm1952suggested2}
	David Bohm.
	\newblock A suggested interpretation of the quantum theory in terms of" hidden"
	variables. ii.
	\newblock \emph{Physical Review}, 85\penalty0 (2):\penalty0 180,
	1952{\natexlab{b}}.
	
	\bibitem[Rosenblatt(1956)]{rosenblatt1956}
	Murray Rosenblatt.
	\newblock Remarks on some nonparametric estimates of a density function.
	\newblock \emph{Ann. Math. Statist.}, 27\penalty0 (3):\penalty0 832--837, 09
	1956.
	\newblock \doi{10.1214/aoms/1177728190}.
	\newblock URL \url{https://doi.org/10.1214/aoms/1177728190}.
	
	\bibitem[Parzen(1962)]{parzen1962}
	Emanuel Parzen.
	\newblock On estimation of a probability density function and mode.
	\newblock \emph{Ann. Math. Statist.}, 33\penalty0 (3):\penalty0 1065--1076, 09
	1962.
	\newblock \doi{10.1214/aoms/1177704472}.
	\newblock URL \url{https://doi.org/10.1214/aoms/1177704472}.
	
	\bibitem[Herrmann et~al.(2017)Herrmann, Hall, Wiseman, and
	Deckert]{miw_kde_gstate}
	Hannes Herrmann, Michael J.~W. Hall, Howard~M. Wiseman, and Dirk~André
	Deckert.
	\newblock Ground states in the many interacting worlds approach, 2017.
	
	\bibitem[Poltavsky and Tkatchenko(2016)]{pimd_tdep}
	Igor Poltavsky and Alexandre Tkatchenko.
	\newblock Modeling quantum nuclei with perturbed path integral molecular
	dynamics.
	\newblock \emph{Chem. Sci.}, 7:\penalty0 1368--1372, 2016.
	\newblock \doi{10.1039/C5SC03443D}.
	\newblock URL \url{http://dx.doi.org/10.1039/C5SC03443D}.
	
	\bibitem[Fletcher(1987)]{fletcher1987practical}
	R~Fletcher.
	\newblock \emph{Practical methods of optimization}.
	\newblock Wiley, Chichester New York, 1987.
	\newblock ISBN 9780471915478.
	
	\bibitem[Craig and Manolopoulos(2004)]{q_rtimecorr}
	Ian~R. Craig and David~E. Manolopoulos.
	\newblock Quantum statistics and classical mechanics: Real time correlation
	functions from ring polymer molecular dynamics.
	\newblock \emph{The Journal of Chemical Physics}, 121\penalty0 (8):\penalty0
	3368--3373, 2004.
	\newblock \doi{10.1063/1.1777575}.
	\newblock URL \url{https://doi.org/10.1063/1.1777575}.
	
	\bibitem[van~der Walt et~al.(2011)van~der Walt, Colbert, and Varoquaux]{numpy}
	Stéfan van~der Walt, S.~Chris Colbert, and Gaël Varoquaux.
	\newblock The numpy array: A structure for efficient numerical computation.
	\newblock \emph{Computing in Science \& Engineering}, 13\penalty0 (2):\penalty0
	22--30, 2011.
	\newblock \doi{10.1109/MCSE.2011.37}.
	\newblock URL \url{http://aip.scitation.org/doi/abs/10.1109/MCSE.2011.37}.
	
	\bibitem[Jones et~al.(2001--)Jones, Oliphant, Peterson, et~al.]{scipy}
	Eric Jones, Travis Oliphant, Pearu Peterson, et~al.
	\newblock {SciPy}: Open source scientific tools for {Python}, 2001--.
	\newblock URL \url{http://www.scipy.org/}.
	\newblock [Online; accessed 2017-05-11].
	
	\bibitem[Pillai et~al.(2012)Pillai, Goglio, and Walker]{mnumerov}
	Mohandas Pillai, Joshua Goglio, and Thad~G. Walker.
	\newblock Matrix numerov method for solving schrödinger’s equation.
	\newblock \emph{American Journal of Physics}, 80\penalty0 (11):\penalty0
	1017--1019, 2012.
	\newblock \doi{http://dx.doi.org/10.1119/1.4748813}.
	\newblock URL
	\url{http://scitation.aip.org/content/aapt/journal/ajp/80/11/10.1119/1.4748813}.
	
	\bibitem[Sheather(2004)]{kde_review}
	Simon~J. Sheather.
	\newblock Density estimation.
	\newblock \emph{Statist. Sci.}, 19\penalty0 (4):\penalty0 588--597, 11 2004.
	\newblock \doi{10.1214/088342304000000297}.
	\newblock URL \url{http://dx.doi.org/10.1214/088342304000000297}.
	
	\bibitem[Bell(1980)]{bell_tunnel}
	Ronald P. Bell
	\newblock \emph{The Tunnel Effect in Chemistry}.
	\newblock Chapman and Hall, New York/London, 1980.

	\bibitem[McKeague et~al.(2016)McKeague, Peköz, and Swan]{stein_miw}
	Ian~W. McKeague, Erol~A. Peköz, and Yvik Swan.
	\newblock Stein's method, many interacting worlds and quantum mechanics, 2016.
	
	\bibitem[Stromberg and Society(1981)]{stromberg1981introduction}
	Karl~Robert Stromberg and American~Mathematical Society.
	\newblock \emph{An introduction to classical real analysis}.
	\newblock Wadsworth International Group Belmont, California, 1981.
	
	\bibitem[Mohebbi and Dehghan(2009)]{MOHEBBI}
	Akbar Mohebbi and Mehdi Dehghan.
	\newblock The use of compact boundary value method for the solution of
	two-dimensional schrödinger equation.
	\newblock \emph{Journal of Computational and Applied Mathematics}, 225\penalty0
	(1):\penalty0 124 -- 134, 2009.
	\newblock ISSN 0377-0427.
	\newblock \doi{http://dx.doi.org/10.1016/j.cam.2008.07.008}.
	\newblock URL
	\url{http://www.sciencedirect.com/science/article/pii/S0377042708003579}.
	
\end{thebibliography}

\end{document}